\documentclass{jps-cp}

\usepackage{graphicx}
\usepackage{color}
\title{Translational Symmetry Broken Magnetization Plateau of the $S=2$ Antiferromagnetic Chain with Anisotropies}

\author{Takaharu~\textsc{Yamada}$^{1}$, Ryosuke~\textsc{Nakanishi}$^{1}$, 
Rito~\textsc{Furuchi}$^1$, Hiroki~\textsc{Nakano}$^1$, Hirono~\textsc{Kaneyasu}$^1$,
Kiyomi~\textsc{Okamoto}$^1$, Takashi~\textsc{Tonegawa}$^{1,2,3}$ and T\^oru~\textsc{Sakai}$^{1,4}$}

\inst{
$^{1}$Graduate School of Science, University of Hyogo, Hyogo 678-1297, Japan \\
$^{2}$Professor Emeritus, Kobe University, Kobe 657-8501, Japan\\
$^{3}$Department of Physics, Graduate School of Science, Osaka Metropolitan University, Sakai, Osaka 599-8531, Japan\\
$^{4}$National Institutes for Quantum Science and Technology (QST), SPring-8, Hyogo 679-5148, Japan
}

\email{takaharu19990128@yahoo.co.jp}

\recdate{July 12, 2022}

\abst{
The magnetization plateau of the $S=2$ antiferromagnetic chain with interaction and single-ion anisotropies is investigated using the 
numerical diagonalization of finite-size clusters and some size scaling analyses. 
The previous level spectroscopy analysis indicated that two different magnetization plateau 
phases appear at half of the saturation magnetization. 
One is due to the large-$D$ mechanism and the other is due to the Haldane one. 
In the present study the phase diagram is extended to wider region of the anisotropies. 
As a result we find another half magnetization plateau phase, where the translational 
symmetry is spontaneously broken .
}

\kword{quantum spin system, magnetization plateau, spin gap, quantum phase transition}

\def\Vec#1{\mbox{\boldmath $#1$}}

\begin{document}
\maketitle

\section{Introduction}

The magnetization plateau is one of interesting topics in the field of the low-temperature physics. 
It was proposed as the Haldane gap induced by the external magnetic field\cite{oya}. 
According to the extended Lieb-Schultz-Mattis theorem\cite{oya,lsm}, the $S=3/2$ antiferromagnetic chain 
was revealed to exhibit the magnetization plateau at $1/3$ of the saturation magnetization, 
even without any translational symmetry breaking. 
The previous numerical diagonalization and some finite-size scaling analyses 
indicated that the $1/3$ magnetization plateau of the $S=3/2$ antiferromagnetic chain would appear 
with some anisotropies\cite{sakai3,kitazawa}. 
The $1/2$ magnetization plateau at half of the saturation magnetization
of the $S=2$ antiferromagnetic chain would also possibly appear 
even without the translational symmetry breaking. 
The recent numerical diagonalization and the level spectroscopy analysis on the 
$S=2$ chain with the single-ion anisotropy $D$ and the $XXZ$ coupling anisotropy $\lambda$ \cite{sakai2019}
indicated that the system exhibited two different $1/2$ magnetization plateau phases; 
one is the Haldane plateau phase and the other is the large-$D$ one. 
The Haldane plateau phase is the symmetry protected topological (SPT) phase\cite{pollmann1,pollmann2}, 
which was also revealed to appear
as the intermediate-$D$ phase in the ground state of the same system 
without magnetic field\cite{tonegawa1,okamoto1,okamoto2,okamoto3,okamoto4}. 
The same numerical diagonalization and the level spectroscopy analysis indicated that 
the biquadratic interaction stabilizes these SPT phases\cite{sakai2022}. 
In these magnetization plateau phases any translational symmetry is not broken. 

In this paper we investigate the same $S=2$ antiferromagnetic chain model 
in wider range of the anisotropies, using some finite-size scaling analyses. 
As a result, we find a new $1/2$ magnetization plateau phase, where the translational 
symmetry is broken. Thus it should be called the N\'eel plateau phase. 
We also obtain the phase diagram at half of the saturation magnetization, 
which includes three plateau phases, which will be shown later.

\section{Model}

The magnetization process 
of the $S=2$ antiferromagnetic Heisenberg chain with the 
exchange and single-ion anisotropies, denoted by $\lambda$ and $D$, 
respectively, is described by 
the Hamiltonian 
\begin{eqnarray}
\label{ham}
&&{\cal H}={\cal H}_0+{\cal H}_Z, \\
&&{\cal H}_0 = \sum _{j=1}^L \left[ S_j^xS_{j+1}^x + S_j^yS_{j+1}^y 
 + \lambda S_j^zS_{j+1}^z \right]
  +D\sum_{j=1}^L (S_j^z)^2, \\
&&{\cal H}_Z =-H\sum _{j=1}^L S_j^z,
\end{eqnarray}
where $S_j^\mu$ denotes the $\mu (=x,y,z)$ component of the $S=2$ operator at the $j$th site,
and $H$ is the external magnetic field.
The exchange interaction constant is set to unity as the unit of energy.
For $L$-site systems, 
the lowest energy of ${\cal H}_0$ in the subspace where 
$\sum _j S_j^z=M$, is denoted as $E(L,M)$. 
The reduced magnetization $m$ is defined as $m=M/M_{\rm s}$, 
where $M_{\rm s}$ denotes the saturation of the magnetization, 
namely $M_{\rm s}=L S$ for the spin-$S$ system. 
$E(L,M)$ is calculated by the Lanczos algorithm under the 
periodic boundary condition ($ \Vec{S}_{L+1}=\Vec{S}_1$) 
and the twisted boundary condition 
($S^{x,y}_{L+1}=-S^{x.y}_1, S^z_{L+1}=S^z_1$), 
up to $L=12$. 
Both boundary conditions are necessary for the level spectroscopy 
analysis. 

\section{Magnetization Plateau}

We focus on the state at $m=1/2$ in the magnetization process 
of the system (\ref{ham}) at $T=0$. 
In this state the magnetization per unit cell is $M/L$=1. 
Thus Oshikawa, Yamanaka and Affleck's theorem\cite{oya} suggests 
that the magnetization plateau possibly occurs
without the spontaneous breaking of the translational symmetry, 
because $S-M/L={\rm integer}$. 
In the previous work\cite{sakai2019}, two different $m=1/2$ magnetization 
plateau phases without the translational symmetry breaking were 
revealed to appear and the phase diagram for $1.0 < \lambda < 2.5$ and 
$0.0 < D <3.0$ was presented. 
The main purpose of this paper is to report a new $m=1/2$ plateau phase 
with the translational symmetry breaking, and present the phase 
diagram in wider region of the anisotropies including the previous 
result. Then we briefly review the previous work about two 
plateau phases without the translational symmetry breaking 
in the next subsection. 

\subsection{Plateau without Symmetry Breaking}

\begin{figure}[tbh]
\bigskip
\centerline{\includegraphics[scale=0.4]{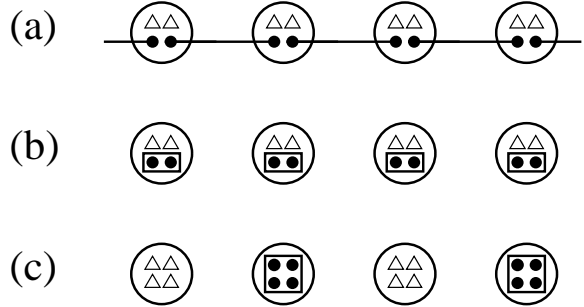}}
\caption{Schematic pictures of (a) the Haldane plateau state, 
        (b) the large-$D$ plateau state, and (c) the N\'eel plateau state.
        An $S=2$ spin consists of four $s=1/2$ spins. 
        Big open circles denote $S=2$ spins and small dots $s=1/2$ spins.
        Open triangles denote $s=1/2$ spins with $s^z=1/2$.
        Solid lines represent singlet dimer pair
        $(1/\sqrt{2})(|\uparrow \downarrow - \downarrow \uparrow \rangle)$
        and rectangles  $(1/\sqrt{2})(|\uparrow \downarrow + \downarrow \uparrow \rangle)$.
        Squares denote the lowest four spin state with $s_{\rm tot}^z = 0$.}
\label{pictures}
\end{figure}

In the previous work, the present system was revealed to exhibit two different 
$1/2$ magnetization plateau phases without the translational symmetry breaking; 
they are the Haldane plateau phase, where the valence bond solid\cite{aklt1,aklt2} 
is realized, and the large-$D$ one. 
Schematic pictures of these two states are shown in Figs. \ref{pictures}(a) and (b).

\begin{figure}[htb]
\bigskip
\centerline{\includegraphics[scale=0.45]{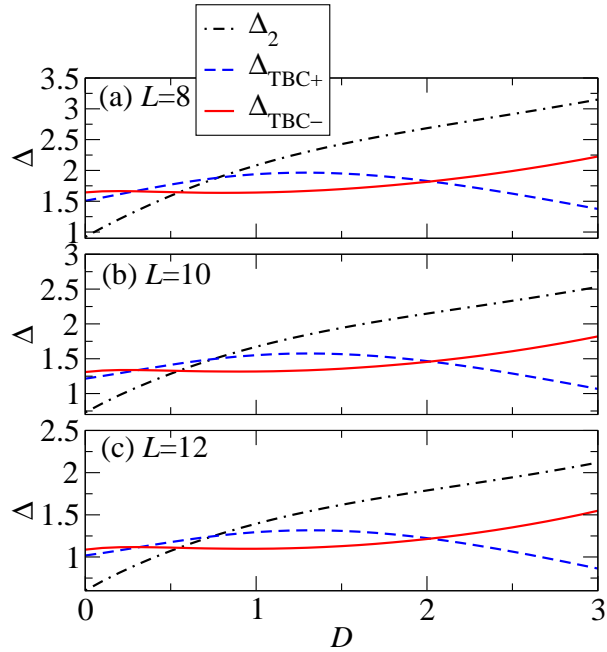}}
\caption{
The $D$ dependence of three gaps $\Delta_2$ (black dashed-dotted line), $\Delta_{\rm TBC+}$
 (blue dashed line), and $\Delta_{\rm TBC-}$ (red solid line) for $\lambda =2.5$ calculated 
for (a) $L=8$, (b) $L=10$ and (c) $L=12$. 
}
\label{LS2}
\end{figure}

These two plateau phases and the gapless Tomonaga-Luttinger liquid (TLL) phase (no plateau phase)
can be distinguished using the level spectroscopy method\cite{kitazawa}. 
According to this analysis, 
we should compare the following three energy gaps; 
\begin{eqnarray}
\label{delta2}
&&\Delta _2 ={E(L,M-2)+E(L,M+2)-2E(L,M) \over 2}, \\
\label{tbc+}
&&\Delta_{\rm TBC+}=E_{\rm TBC +}(L,M)-E(L,M), \\
\label{tbc-}
&&\Delta_{\rm TBC-}=E_{\rm TBC -}(L,M)-E(L,M),
\end{eqnarray}
where $E_{\rm TBC+}(L,M)$ ($E_{\rm TBC-}(L,M)$) is the energy of the 
lowest state with the even parity (odd parity) with respect to the space inversion 
at the twisted bond under the twisted boundary condition, 
and other energies are under the periodic 
boundary condition. 
The level spectroscopy method indicates that the smallest gap 
among these three gaps for $M=L=M_{\rm s}/2$ determines the phase 
at $m=1/2$. 
Namely, $\Delta_2$, $\Delta_{\rm TBC+}$ and $\Delta _{\rm TBC-}$ 
correspond to the TLL, large-$D$-plateau and Haldane-plateau phases, 
respectively.

The $D$ dependence of the three gaps calculated for $L=8, 10$, and $12$ 
is plotted for $\lambda =2.5$ in Figs. \ref{LS2}(a), (b), and (c), respectively. 
These figures indicate that the system is in the no-plateau phase for small $D$, 
in the Haldane plateau phase for intermediate $D$, 
and in the large-$D$ plateau phase for large $D$. 
The system size dependence of the boundary is predicted to 
proportional to $1/L^2$. 

Thus assuming the finite-size correction proportional to $1/L^2$, 
we estimate each phase boundary in the infinite length limit 
from the intersections of the gaps for $L=$8, 10 and 12. 




\subsection{Translational Symmetry Broken Plateau}

\begin{figure}[tbh]
\bigskip
\bigskip
\centerline{\includegraphics[scale=0.35]{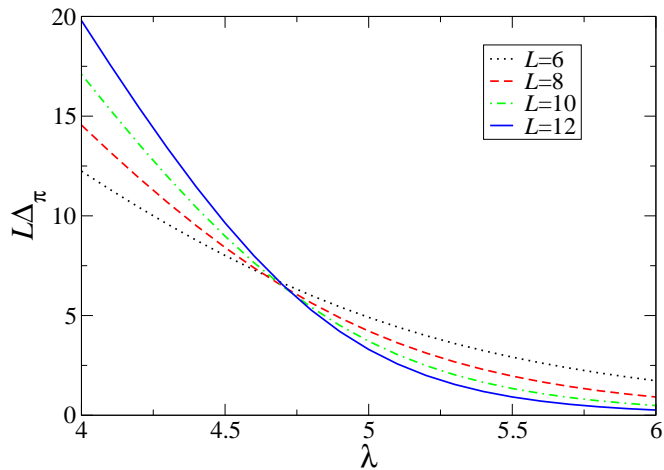}}
\caption{
The scaled gap $L\Delta _{\pi}$ for $D=2.0$ is plotted versus $\lambda$ 
for $L=$6, 8, 10 and 12.
}
\label{prg}
\end{figure}

\begin{figure}[h]
   \begin{minipage}{.48\linewidth}
       \centerline{\includegraphics[scale=0.22]{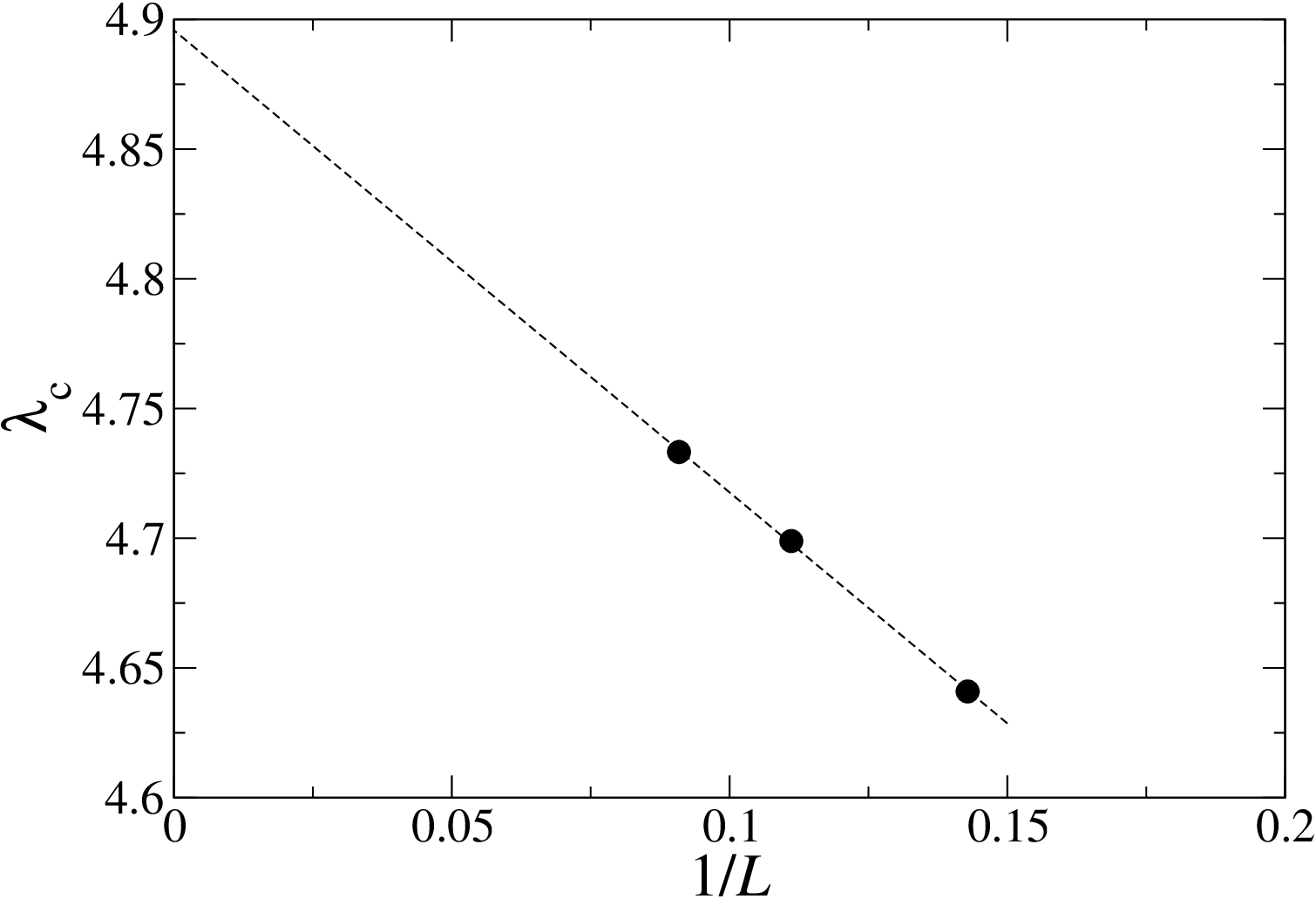}}
       \caption{
       The size-dependent fixed point $\lambda_{\rm c}(L)$ for $L=$7, 9, 11 is 
       plotted versus $1/L$ for $D=2.0$.  
       Assuming the size correction proportional to $1/L$, 
       $\lambda_{\rm c}$ in the infinite length limit is estimated as 
       $\lambda_{\rm c} = 4.896 \pm 0.004$.
       }
       \label{gaisoD2}
   \end{minipage}
    \hspace{0.04\columnwidth}
   \begin{minipage}{.48\linewidth}
       \centerline{\includegraphics[scale=0.22]{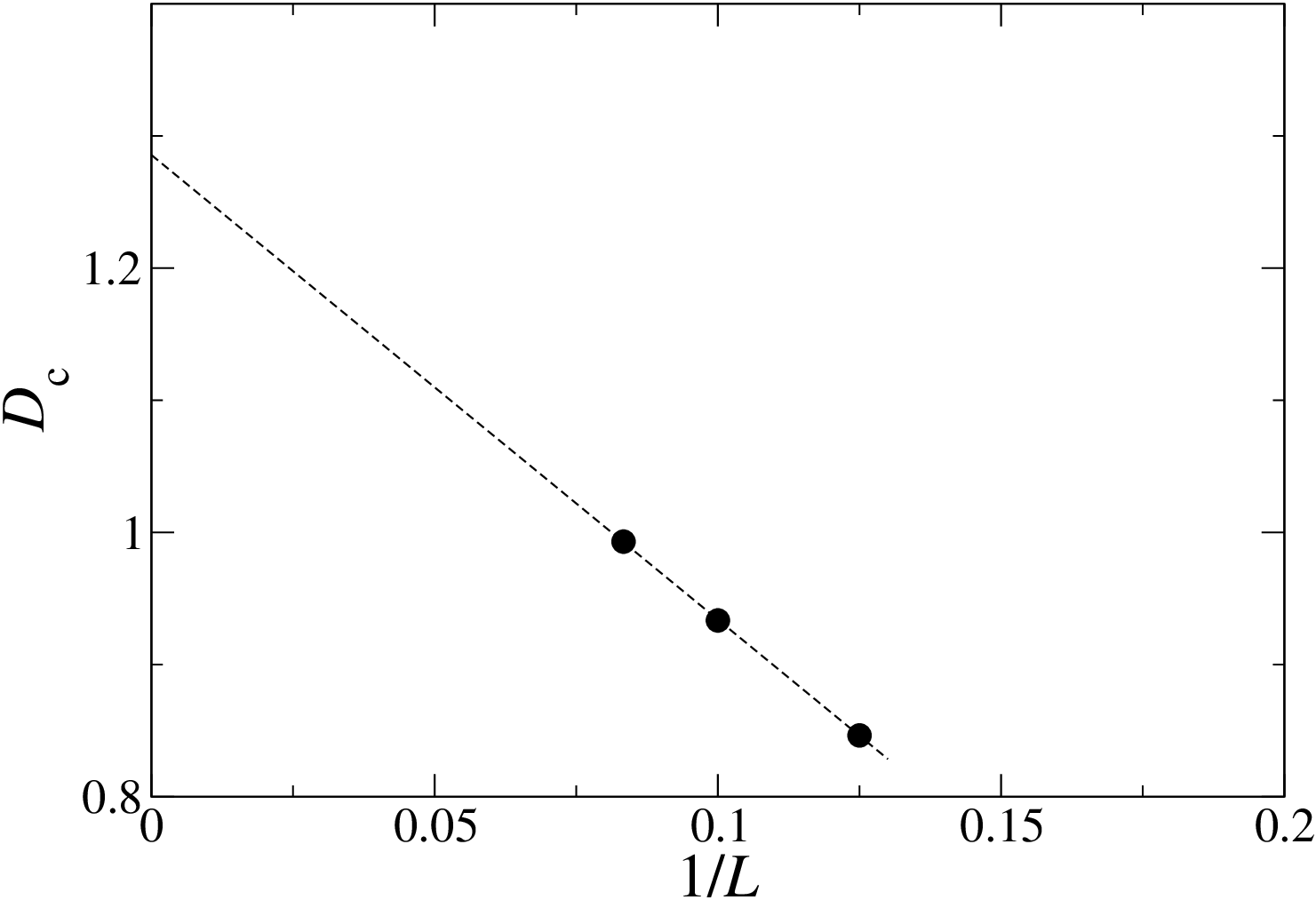}}
       \caption{
       The boundary of the missing region $D_{\rm c}(L)$ for $\lambda=5.0$ 
       is plotted versus $1/L$.
       Assuming the size correction proportional to $1/L$, 
       $D_{\rm c}$ in the infinite length limit is estimated as 
       $D_{\rm c}=1.286 \pm 0.003$. 
       }
       \label{missing}
   \end{minipage}
\end{figure}

If the Ising-like anisotropy $\lambda$ is sufficiently large, 
the system is expected to exhibit another $m=1/2$ magnetization plateau 
phase, where the translational symmetry is spontaneously broken like the 
N\'eel order, namely $|\cdots ,2,0,2,0,\cdots \rangle$. 
It should be called the N\'eel plateau phase. 
The schematic picture of the N\'eel plateau state
is shown in Fig. \ref{pictures}(c).
In this phase the excitation with the momentum $k=\pi$ would 
be degenerate to the lowest state with $k=0$. 
Thus the phenomenological renormalization is a good method
to detect the quantum phase transition to the N\'eel plateau phase. 
The lowest excitation gap with $k=\pi$ in the subspace $m=1/2$ 
is denoted as $\Delta_{\pi}$. 
The size-dependent fixed point $\lambda _c(L+1)$ is determined by the equation 
\begin{eqnarray}
L\Delta_{\pi}(L,\lambda)=(L+2)\Delta_{\pi}(L+2,\lambda).
\label{prg-equation}
\end{eqnarray}
The scaled gap $L\Delta _{\pi}$ for $D=2.0$ is plotted versus $\lambda$ 
for $L=$6, 8, 10 and 12 in Fig. \ref{prg}. 
The size-dependent fixed point $\lambda _c(L)$ for $L=$7, 9, 11 is 
plotted versus $1/L$ for $D=2.0$ in Fig. \ref{gaisoD2}. 
$\lambda_{\rm c}$ in the infinite length limit is estimated as 
$\lambda_{\rm c}=4.896 \pm 0.004$. 
The boundary of the N\'eel plateau phase is estimated by this method.

\subsection{Phase diagram}

Apart from the gapless (no plateau) and the magnetization plateau phases, 
there is a parameter region where the $m=1/2$ magnetization is not 
realized due to the magnetization jump. 
%
%
We think this jump corresponds to the spin flop transition \cite{sakai1999-1,sakai1999-2}
where the N\'eel order parallel to the 
external field changes to the quasi-N\'eel order perpendicular to it. 
We can find the boundary of this missing region by checking the magnetization curve.
Namely, if the $m=1/2$ magnetization is included in the magnetization jump,
the system is in the missing region.
The boundary of the missing region $D_{\rm c}$ for $\lambda=5.0$ 
is plotted versus $1/L$ in Fig. \ref{missing}. 
Assuming the size correction proportional to $1/L$, 
$D_{\rm c}$ in the infinite length limit is estimated as 
$D_{\rm c}=1.286 \pm 0.003$. 
The boundary of the missing region is determined by this method. 

\begin{figure}[htb]
\bigskip
\centerline{\includegraphics[scale=0.4]{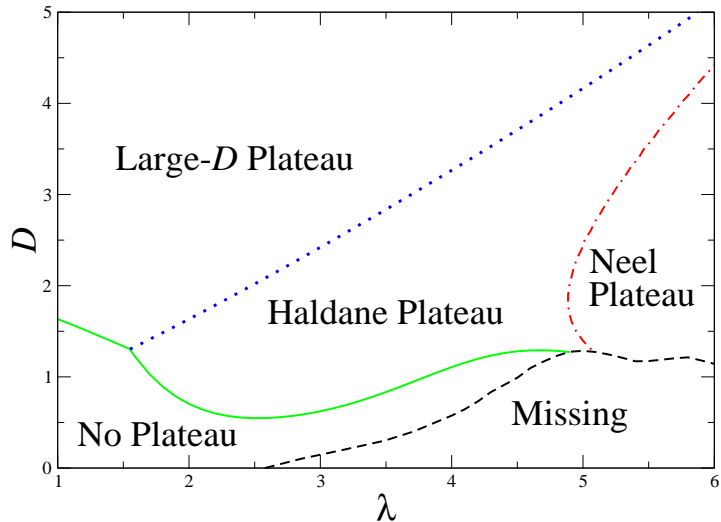}}
\caption{
Ground state phase diagram at $m=1/2$ of the present model. 
It includes the gapless (no plateau), the Haldane, the large-$D$, the N\'eel plateau phases, 
and the missing region. 
}
\label{phase}
\end{figure}

Finally we present the phase diagram at half of the saturation magnetization 
with respect to the anisotropies $\lambda$ and $D$ shown in Fig. \ref{phase}. 
It includes the gapless (no plateau), the Haldane, the large-$D$, the N\'eel plateau phases, 
and the missing region. 
In the N\'eel plateau phase the translational symmetry is broken.

\section{Summary}

The $m=1/2$ magnetization state of the $S=2$ antiferromagnetic chain with the anisotropies $\lambda$ and $D$ 
is investigated using the numerical diagonalization and some finite-size scaling analyses. 
As a result we found a new $1/2$ magnetization plateau phase with the translational symmetry breaking. 
The phase diagram at half of the saturation magnetization is presented..

\section*{Acknowledgment}
This work has been partly supported by JSPS KAKENHI, Grant Numbers 16K05419, 
16H01080 (J-Physics), 18H04330 (J-Physics), JP20K03866, and JP20H05274. 
We also thank the Supercomputer Center, Institute for Solid State Physics,
University of Tokyo and the Computer Room, Yukawa Institute for Theoretical
Physics, Kyoto University for computational facilities.
We have also used the computational resources
of the supercomputer Fugaku provided by the RIKEN
through the HPCI System Research projects (Project ID:
hp200173, hp210068, hp210127, hp210201, and hp220043).

\end{document}